\documentclass[prd,preprint,superscriptaddress,a4paper,showpacs,showkeys,11pt,nofootinbib]{revtex4}
\usepackage{graphicx}
\usepackage{graphics}
\usepackage{dcolumn}
\usepackage{bm}
\usepackage{multirow}
\usepackage{tabularx}
\usepackage{hyperref}
\usepackage{commath}
\usepackage{epstopdf}
\usepackage[T1]{fontenc}
\usepackage[latin9]{inputenc}
\usepackage{geometry}
\usepackage{soul}
\usepackage{amsmath,amssymb,amsfonts}

\usepackage{color}

\def\be{\begin{equation}}
\def\ee{\end{equation}}

\providecommand{\ee}{e$^+$e$^-$}

\newcommand{\gaga}{$\gamma\gamma$}
\newcommand{\gaP}{$\gamma\mathbb P$}
\newcommand{\PP}{$\mathbb P\mathbb P$}

\makeatother

\newcommand{\pom}{\tt I\! P}

\begin{document}

%
%
\title{Top quark pair production in the exclusive processes at LHC}

\author{Victor P. Gon\c calves}
\email[]{barros@ufpel.edu.br}
\affiliation{Instituto de F\'{\i}sica e Matem\'atica, Universidade Federal de
Pelotas (UFPel),\\
Caixa Postal 354, CEP 96010-090, Pelotas, RS, Brazil}

\author{Daniel E. Martins}
\email[]{dan.ernani@gmail.com}
\affiliation{Instituto de F\'isica, Universidade Federal do Rio de Janeiro (UFRJ), 
Caixa Postal 68528, CEP 21941-972, Rio de Janeiro, RJ, Brazil}

\author{Murilo S. Rangel}
\email[]{murilo.rangel@ufrj.br}
\affiliation{Instituto de F\'isica, Universidade Federal do Rio de Janeiro (UFRJ), 
Caixa Postal 68528, CEP 21941-972, Rio de Janeiro, RJ, Brazil}

\author{Marek Tasevsky}
\email[]{Marek.Tasevsky@cern.ch}
\affiliation{Institute of Physics of the Czech Academy of Sciences, 
Na Slovance 2, 18221 Prague 8, Czech Republic}


\begin{abstract}
  We analyze the LHC prospects for measurements of the $t\bar{t}$ pair produced
  exclusively in photon-photon or semi-exclusively in photon-Pomeron and
  Pomeron-Pomeron processes using protons tagged in forward proton detectors on
  both sides of the interaction point. These processes are interesting from the
  point of view of a possible measurement of the top quark mass and constraining
  models used in Beyond Standard Model physics. Focusing on the semi-leptonic
  channel, $t\bar{t}\rightarrow jjbl\nu_l\bar{b}$, making use of the exclusive
  nature of the final state, together with the use of timing information
  provided by forward proton detectors, relevant exclusive and inclusive
  backgrounds are studied in detail for different luminosity (or pile-up)
  scenarios and found to be important for further considerations.
  {While good prospects are found for observing the signal, the top quark
    mass measurement turns out not to be competitive with measurements in
    inclusive channels.}
\end{abstract}




\maketitle
\section{Introduction}\label{sec:intro}
One of the more important processes to study the perturbative Quantum
Chromodynamics (pQCD) is the heavy quark production in hadronic collisions (for
review, see e.g. Ref~\cite{review_hq}). Such process is expected to improve
the description of the measured data by pQCD at high energies and it is also an
important background for analyses searching for signals of Beyond Standard
Model (BSM) processes. In other words, the analysis of the top quark production
allows us to constrain input parameters for pQCD predictions and probe different
BSM scenarios \cite{Husemann:2017eka,Fayazbakhsh:2015xba,Howarth:2020uaa}. This expectation is directly related to the fact that the top
quark couples to all gauge bosons and the Higgs boson, which implies that the
top production is very sensitive to the presence of BSM phenomena. In addition,
recent experimental results for the inclusive top pair production have
demonstrated that this process can be used to measure the top mass in a
well-defined scheme with a high accuracy 
\cite{Aad:2019mkw,Aad:2019ntk,Aad:2019hzw,Sirunyan:2018wem,Sirunyan:2017mzl,Sirunyan:2018goh}. These results motivate the study of
the top pair production in diffractive processes, where the final state is
cleaner in comparison to the typical inclusive one, where both incident protons
fragment and a large number of particles are produced in addition to the top
pair (For  reviews about diffraction see e.g. Ref. \cite{pomeron}). In our analysis we will focus on the top pair production in photon- and
pomeron-induced processes at the center of system energy, $\sqrt s$, of 13~TeV,
where both incident protons remain intact in the
final state. In principle, such events can be collected using the forward proton
detectors (FPD) such as the ATLAS Forward Proton detector (AFP)
\cite{Adamczyk:2015cjy,Tasevsky:2015xya} and Precision Proton Spectrometer
(CT-PPS) \cite{Albrow:2014lrm} that are installed symmetrically around
the interaction point at a distance of roughly 210~m from the interaction point. 
{We will restrict our analysis to Standard Model subprocesses and postpone the study of the impact of new physics on the diffractive  $t\bar{t}$ production for a future publication.}

\begin{figure}[t]
\begin{center}
\begin{tabular}{ccc}
\scalebox{0.38}{\includegraphics{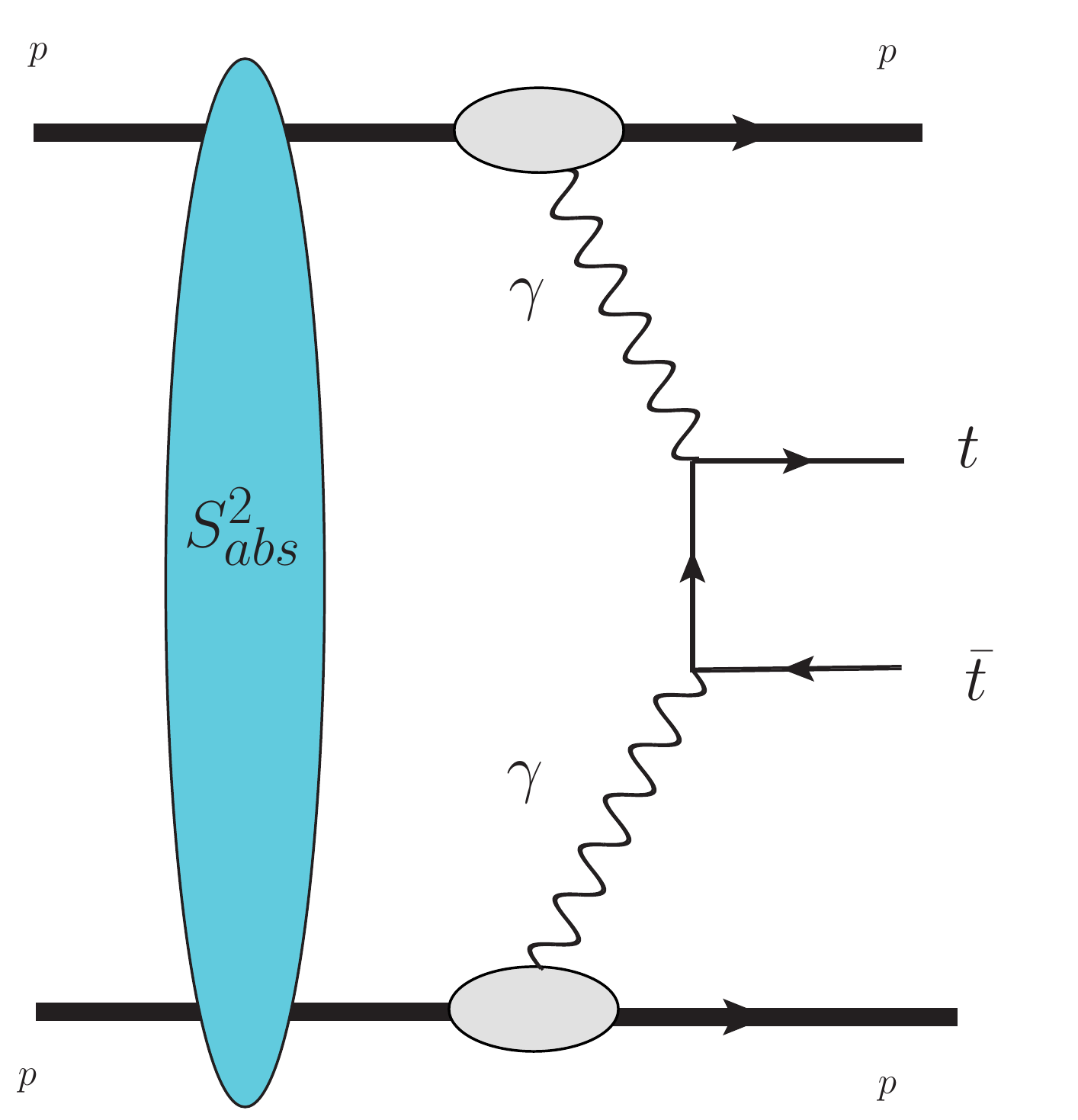}}&\scalebox{0.28}{\includegraphics{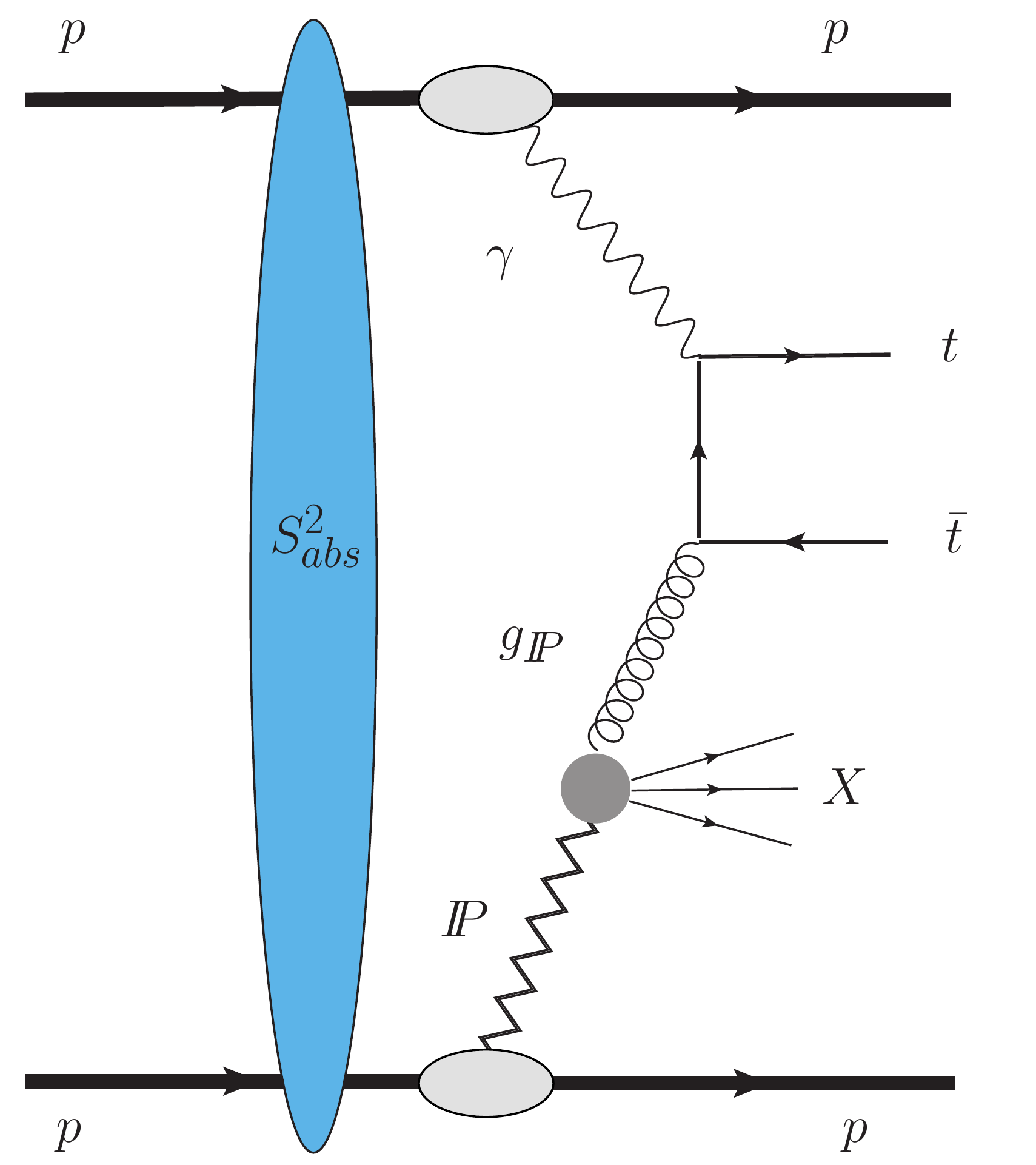}} &\scalebox{0.28}{\includegraphics{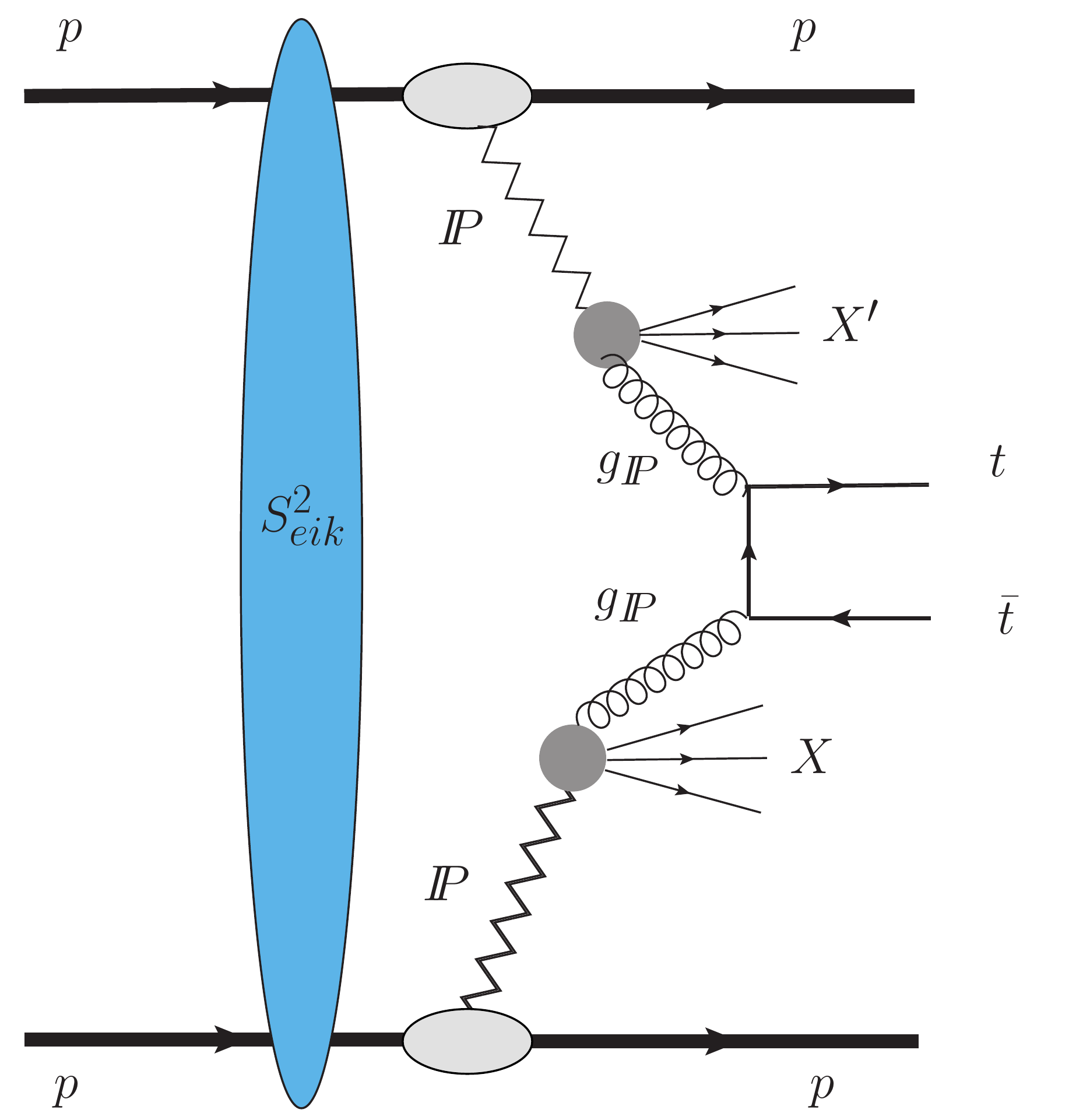}}\\
(a) & (b) & (c) \\
\end{tabular}
\caption{Top pair production in (a) photon--photon, (b) photon--pomeron and
  (c) pomeron--pomeron interactions in $pp$ collisions.}
\label{Fig:diagram}
\end{center}
\end{figure}

At high energies the top pair production in hadronic collisions is dominated by
gluon--gluon interactions provided that the incident hadrons break up. However,
a top
pair can also be generated in photon--photon (Fig.~\ref{Fig:diagram} (a)), 
photon--pomeron (Fig.~\ref{Fig:diagram} (b)) and  pomeron--pomeron (Fig.~\ref{Fig:diagram} (c))  interactions. Since the photon and pomeron are color-singlet
objects, these processes are characterized by the presence of two regions devoid
of hadronic activity, called rapidity gaps, separating the intact very forward
protons from the central massive object. Moreover, the process (a) is a typical
example of exclusive process, where nothing else is produced except the leading
hadrons and the central object. In contrast, if we assume that the Pomeron has
a partonic structure \cite{IS}, then in processes (b) and (c) rapidity gaps can be filled
by particles from fragmenting Pomeron remnants.
The large invariant mass of the produced system implies that the intact protons
in the final state can be tagged by FPDs.
Consequently, such events can, in principle, be separated and be used to
improve our understanding of the top quark production. 
Our goal in this paper is to perform a detailed analysis of the top pair
production considering the processes shown in Fig.~\ref{Fig:diagram} and
present expected event yields that take into account the current detector
acceptances and pile-up effects expected for the next run of LHC.
In order to obtain realistic predictions for the top pair production in photon-
and pomeron-induced interactions and to be able to include experimental cuts in
the calculations, the treatment of these processes in a Monte Carlo
simulation is indispensable. Some years ago, the Forward Physics Monte Carlo
(FPMC) \cite{fpmc} was generalized in order to simulate the central particle
production with one or two leading protons and some hard scale in the event.  
In this paper we use this generator to estimate the top pair production at
the LHC. We try to give realistic estimates of the event yields for the signal
as well as backgrounds and comment also on possibilities to extract information
about the top quark mass using FPDs.

This paper is organized as follows. In the next section we present a brief
review of the formalism for the top pair production in photon- and
pomeron-induced interactions in $pp$ collisions. In Section~\ref{sec:exp}
we discuss details of the selection of events and cuts implemented in our
analysis, concentrating on collisions at $\sqrt s = 13$~TeV. In
Section~\ref{sec:results} we present our predictions for the
invariant mass and transverse momentum distributions as well as for the total
cross sections for the top pair production in \gaga, \gaP\ and \PP\
interactions. Finally, in Section~\ref{sec:sum} we summarize our main
conclusions.

\section{Formalism}\label{sec:form}
An ultra relativistic charged hadron (proton or nucleus) gives rise to strong
electromagnetic fields, such that the photon stemming from the electromagnetic
field of one of the two colliding hadrons can interact with one photon of the
other hadron (photon--photon process) or can interact directly with the other
hadron (photon--hadron process) \cite{upc,epa}.  The total cross section for the
photon-induced interactions can be factorized in terms of the equivalent flux of
photons into the hadron projectiles and the photon--photon or photon--target
production cross section. In particular, the top pair production in photon--photon interactions, represented in Fig.~\ref{Fig:diagram} (a), is described by
\begin{eqnarray}
\sigma(h_1 h_2 \rightarrow h_1 \otimes t\bar{t} \otimes h_2) = \int dx_1 \int dx_2 \, \gamma_1(x_1) \cdot \gamma_2(x_2) \cdot \hat{\sigma}(\gamma \gamma \rightarrow t \bar{t}) \,\,,
\label{fotfot}
\end{eqnarray}
where $\otimes$ represents the presence of a rapidity gap in the
final state, $x$ is the fraction of the hadron energy carried by the photon
and $\gamma(x)$ is the equivalent photon distribution of the hadron. 
The general expression for the photon flux of the proton is given by \cite{kniehl} 
\begin{eqnarray}
\gamma (x) = - \frac{\alpha}{2\pi} \int_{-\infty}^{-\frac{m^2x^2}{1-x}} \frac{dt}{t}\left\{\left[2\left(\frac{1}{x}-1\right) + \frac{2m^2x}{t}\right]H_1(t) + xG_M^2(t)\right\}\,\,,
\label{elastic}
\end{eqnarray}
where $t = q^2$ is the momentum transfer squared of the photon,
\begin{eqnarray}
 H_1(t) \equiv \frac{G_E^2(t) + \tau G_M^2(t) }{1 + \tau}
\end{eqnarray}
with $\tau \equiv -t/m^2$, $m$ being the nucleon mass, and where $G_E$ and
$G_M$ are the Sachs elastic form factors. In what follows we will use the photon
flux derived in Ref.~\cite{epa}, where an analytical expression is presented (for a recent study of the $t\bar{t}$ production in $\gamma \gamma$ interactions see, e.g. Ref. \cite{Luszczak:2018dfi}).

The photon--hadron processes can be classified as inclusive or diffractive,
depending if the proton breaks up or remains intact, respectively. The inclusive $t\bar{t}$ production by inclusive $\gamma p$ interactions in $pp$ collisions at the LHC was analyzed in Ref. \cite{Goncalves:2013oga}. In our
study we are interested in diffractive photon--hadron case, with the diffractive
interaction being described by a Pomeron exchange. The cross section for the top
pair production in a photon--pomeron interaction, represented in
Fig.~\ref{Fig:diagram} (b), is given by
\begin{eqnarray}
\sigma(h_1 h_2 \rightarrow h_1 \otimes t\bar{t} X \otimes h_2 ) = \int dx_1 \int dx_2 \, [g^D_1(x_1,\mu^2) \cdot \gamma_2(x_2) + \gamma_1(x_1) \cdot g^D_2(x_2,\mu^2)] \cdot \hat{\sigma}(\gamma g \rightarrow t \bar{t}) \,\,,
\label{pompho}
\end{eqnarray}
where  $g^D (x,\mu^2)$ is the diffractive gluon distribution, whose evolution is
described by the DGLAP evolution equations and is determined from events with a
rapidity gap or intact proton, mainly at HERA \cite{pomeron}. Similarly, the top pair can be
produced by pomeron--pomeron interactions, as represented in Fig.~\ref{Fig:diagram} (c), with the cross section being given by
\begin{eqnarray}
\sigma(h_1 h_2 \rightarrow h_1 \otimes X t\bar{t} X^{\prime} \otimes h_2) = \int dx_1 \int dx_2 \, g^D_1(x_1,\mu^2) \cdot g^D_2(x_2,\mu^2) \cdot \hat{\sigma}(g g \rightarrow t\bar{t}) \,\,.
\label{pompom}
\end{eqnarray}
In the resolved Pomeron model \cite{IS}, the diffractive gluon distribution in the proton,
$g^D (x,\mu^2)$, is defined as a convolution of the Pomeron flux emitted by
proton, $f_{I\!\!P}(x_{I\!\!P})$, and the gluon distribution in Pomeron,
$g_{I\!\!P}(\beta, \mu^2)$,  where $\beta$ is the momentum fraction carried by
the struck parton inside the Pomeron. 
The Pomeron flux is given by $f_{I\!\!P}(x_{I\!\!P})= \int_{t_{min}}^{t_{max}} dt f_{\pom/p}(x_{{I\!\!P}}, t)$, where $f_{\pom/p}(x_{\pom}, t) = A_{\pom} \cdot \frac{e^{B_{\pom} t}}{x_{\pom}^{2\alpha_{\pom} (t)-1}}$ and $t_{min}$, $t_{max}$ are kinematic boundaries.
The Pomeron flux factor is motivated by Regge theory, where the Pomeron
trajectory is assumed to be linear, $\alpha_{\pom} (t)= \alpha_{\pom} (0) + \alpha_{\pom}^\prime t$, and the parameters $B_{\pom}$, $\alpha_{\pom}^\prime$ and their
uncertainties are obtained from fits to H1 data \cite{H1diff}. 
The diffractive gluon distribution is then given by
\begin{eqnarray}
{ g^D(x,\mu^2)}=\int dx_{I\!\!P}d\beta \delta (x-x_{I\!\!P}\beta)f_{I\!\!P}(x_{I\!\!P})g_{I\!\!P}(\beta, \mu^2)={ \int_x^1 \frac{dx_{I\!\!P}}{x_{I\!\!P}} f_{I\!\!P}(x_{I\!\!P}) g_{I\!\!P}\left(\frac{x}{x_{I\!\!P}}, \mu^2\right)}
\end{eqnarray}
Similar definition can be established for the diffractive quark distributions. 
In our analysis we will include the quark contributions for the top pair
production, associated e.g. to the $q\bar{q} \rightarrow t \bar{t}$ subprocess,
and the diffractive parton distribution will be described by the
parameterization obtained by the H1 Collaboration at DESY-HERA, denoted as fit
A in Ref.~\cite{H1diff}. For a similar analysis for the charm and bottom production in \gaP\ and \PP\ interactions see, e.g., Refs. \cite{victor, antoni2,potterat}.  
{{Our predictions for the $t\bar{t}$ production could be potentially sensitive to the Pomeron gluon distribution at large values of $\beta$ and $\mu^2$, beyond the kinematical range probed by HERA. The behavior of $g_{\pom}$ in this region is driven by the DGLAP evolution equations. As the inclusive $t\bar{t}$ data are quite well described by predictions derived using these equations (see e.g. Ref.~\cite{Aad:2019hzw}) and the results presented in Ref.~\cite{Guzey} indicate that the uncertainty on $g_{\pom}$ is small in that kinematical range, we expect, as well, a small impact on our predictions associated to the choice of the diffractive parton distribution.}}

One important open question is the treatment of additional soft interactions
between incident protons which leads to an extra production of particles that
destroy the rapidity gaps in the final state \cite{bjorken}. As these effects
have a nonperturbative nature, they are difficult to treat and their magnitude is
strongly model dependent (for reviews see Refs.~\cite{durham,telaviv,sgap4}). For
photon--photon and photon--pomeron interactions, the contribution of the soft
interactions, represented by the factor $S^2_{abs}$ in Figs.~\ref{Fig:diagram}
(a) and (b), is expected to be small due to the long range of the
electromagnetic interaction. As a consequence, in what follows we will assume
that $S^2_{abs} = 1$. In contrast, for pomeron--pomeron interactions, the impact
of soft interactions is non-negligible, implying the violation of the QCD hard
scattering factorization theorem for diffraction in $pp$ collisions \cite{collins}.  Assuming that the hard process occurs on 
a short enough timescale such that the physics that generate the additional particles can be factorized, the
inclusion of these additional absorption effects can be parameterized in terms of
an average rapidity gap survival probability, $S^2_{eik}$. Such quantity corresponds to
the probability of the scattered proton not to dissociate due to the secondary
interactions. The gap survival probability has been calculated considering different
approaches giving distinct predictions (see, e.g. Ref.~\cite{review_martin}).
As in previous studies for single and double diffractive production \cite{potterat,antoni,antoni2,dimuons,palota} we also follow this simplified approach
assuming $S^2_{eik} = 0.03$ for pomeron--pomeron interactions as predicted in Ref.~\cite{KMR}. However, it is
important to emphasize that the magnitude of the rapidity gap survival probability is still an open
question. 


\section{Experimental procedure}\label{sec:exp}
In this section, we explain the cuts used to discriminate signal from the background sources
considering the pile-up effect. 
We provide results for four
working points regarding the instantaneous luminosity (and hence the average
amount of pile-up interaction per bunch crossing, $\langle \mu \rangle$) and
assumed integrated luminosity in each of the working points. We consider these
values of $\langle \mu \rangle$ and corresponding integrated luminosity:
$\langle \mu \rangle$ = 0 (1~fb$^{-1}$), 5 (10~fb$^{-1}$), 10 (30~fb$^{-1}$) and
50 (300~fb$^{-1}$).
The separation of the (semi)-exclusive $t\bar{t}$ signal from backgrounds at
13~TeV collisions proceeds in two steps: first we select the central system as
in the inclusive processes, then we apply exclusivity criteria. We select the
so-called semi-leptonic $t\bar{t}$ decays:
$t\bar{t}\rightarrow jjbl\nu_l\bar{b}$,
(one top quark decaying hadronically into two light quarks and a b-quark, the
other into a b-quark and a W boson which then decays leptonically into a lepton
and neutrino). In inclusive interactions, the semi-leptonic channel was shown to
give an optimum signal yield while keeping the purity of the signal still
reasonably high. For example for the ATLAS semi-leptonic channel at 8~TeV
\cite{Aad:2015mbv} the total background contamination was kept at a level of
10\%, with single-top and W+jets background processes contributing most, both by
about 3.5--4\%, the rest coming from Z+jets and multi-jet backgrounds. Thus, by
using the same cuts as in \cite{Aad:2015mbv}, we ensure that the backgrounds
from the four inclusive processes above are kept reasonably low even in the
presence of pile-up.
In the second step, we make use of the (semi)-exclusive nature of our signal
and apply exclusivity cuts which basically means requiring both forward protons
to be tagged in FPDs and requiring large rapidity gaps. They are very powerful
in reducing the inclusive backgrounds but the price to pay is a rather low
signal cross section. In the presence of
pile-up, rapidity gaps are filled by soft particles, so we can only require all
final state objects to be well-isolated from each other and not to be
accompanied by large numbers of particles. By having protons tagged on both
sides of FPD, we can also utilize time-of-flight (ToF) detectors which are very
useful to suppress a combinatorial background coming from pile-up, see e.g.
\cite{Harland-Lang:2018hmi,Tasevsky:2014cpa,ToFperformance}. 

The signal processes (the $t\bar{t}$ production in \gaga, \gaP\ and \PP\
processes) and the exclusive WW production, $\gamma\gamma\rightarrow WW$, for
background are
analyzed using FPMC \cite{fpmc}, while the background from photoproduction of
the single top, $\gamma I\!\!P \rightarrow Wt$, is studied using MadGraph~5
\cite{madgraph} and PYTHIA~8~\cite{pythia} (for previous studies in inclusive processes see, e.g. Refs. \cite{deFavereaudeJeneret:2008hf,Sun:2014qoa,Sun:2014exa}). The main background, namely the
inclusive $t\bar{t}$ production with pile-up, is generated using MadGraph~5 and
PYTHIA~8. While the signal processes and the inclusive background
are studied at detector level, the contamination by the exclusive backgrounds
is estimated at generator level. Detector effects and pile-up mixing are
incorporated using Delphes \cite{delphes3} with an input card with ATLAS
detector specifications. For both the exclusive and inclusive $t\bar{t}$
processes, the mass of the top quark is set to the value of
174.0~GeV. For FPDs we assume fully efficient reconstruction in the range
$0.015 < \xi_{1,2} < 0.15$, where $\xi_{1,2} = 1 - p_{z 1,2}/E_{\rm beam}$ is the
fractional proton momentum loss on either side of the interaction point
(side 1 or 2). This, in principle, allows one to measure masses of the central
system by the missing mass method, $M\!=\!\sqrt{\xi_1\xi_2s}$, starting from
about 200~GeV. 
Large samples of the aforementioned processes have been generated corresponding
to luminosities that sufficiently exceed those delivered or expected to be
delivered by LHC in the future. The cuts used for these generations are looser
than those used in the analysis to account for detector effects (e.g.
$0.005 < \xi_{1,2} < 0.20$). 

\subsection{Backgrounds}
Considering the signal to be a sum of contributions from all three signal
processes ($t\bar{t}$ produced in \gaga, \gaP\ and \PP\ processes), we can
divide relevant backgrounds to {\it irreducible} (where there are two intact
protons on both sides) and {\it reducible} (where the hard-scale process itself
does not provide intact protons but when overlaid with pile-up interactions,
it forms a dangerous background). We study in detail two relevant irreducible
backgrounds, namely the photoproduction of the single top quark in diffractive interactions,
$\gamma I\!\!P\rightarrow Wt$, and the exclusive WW production,
$\gamma\gamma\rightarrow WW$. Due to their relatively low cross sections, it is
sufficient to study them at generator level. The cross section of the inclusive
production of the $t\bar{t}$ pair is much higher and therefore it is studied
at detector level including pile-up.

\subsection{Jets}
Jets are reconstructed from particles at generator level or from tracks at detector
level using the Anti-kt
algorithm with a radius $R = 0.4$, incorporated inside the FastJet package
\cite{fastjet:2012}. We require at least four jets in total, out of which at
least two to be b-tagged, all having transverse momenta $E_{\rm T,jet} > 25$~GeV
and pseudorapidity $|\eta_{\rm jet}| < 2.5$.
At generator level, a jet is considered to be b-tagged if a B-hadron is found
inside a cone of $R = 0.4$ from the jet axis.
The b-tagging at detector level is based on finding a parton inside a cone of
$R = 0.4$ from the jet axis. If the parton is a b-quark, the b-tag efficiency
formula is applied to get a probability to find a b-tagged jet. If the parton
is a light quark (or gluon), the misidentification rate for light-quark (or
gluon) jets is applied to get a probability to misidentify a light-quark (or
gluon) jet as the b-quark jet. All the b-tag efficiency and misidentification
rates are given as functions of jet $E_{\rm T}$ and $\eta$. The efficiency
formulas are provided inside the Delphes input card and requiring at least two
jets to be b-tagged means that at least two jets have the b-tagging efficiency
greater than 70\%.

\subsection{Leptons}
In the analysis, we require at least one isolated lepton to be found, either
an electron or muon, possibly coming also from $\tau$ decays, with
$E_{\rm l} > 25$~GeV and $|\eta_{\rm l}| < 2.5$. The lepton is considered to be
isolated if the radius
difference between the jet axis and the lepton is $\Delta R_{\rm l,j} > 0.2$. The
electron and muon reconstruction efficiencies as specified in the Delphes input
card are applied as functions of $E_{\rm T}$ and $\eta$ at both generator
and detector level. The leptons are
represented by corresponding particles at generator level and corresponding
objects at detector level (electrons/positrons as clusters in calorimeters and
muons as combined objects at muon spectrometers).

\subsection{Exclusivity cuts}
The exclusive or semi-exclusive nature of our signal enables us to use two
powerful cuts: we require i) both intact protons to be detected by FPDs (the so
called "double-tag") and ii) large rapidity gaps.
Applying the first cut means accounting for the FPD acceptance which is in
general a function of $\xi$ and $p_{\rm T}$ of the intact proton. For simplicity
and not loosing much of generality, we assume that the acceptance is close to
100\% in the range of $0.015 < \xi_{1,2} < 0.15$, where $\xi_{1,2}$ values are
obtained from protons at generator level. The inclusive backgrounds
would naturally not survive such a cut but the combinatorial background from
e.g. on average 50 pile-up interactions in one event gives a non-negligible
probability to see double-tagged events. Most often they come from two soft
Single-diffractive (SD) events each providing a proton in the FPD acceptance on
one side from the interaction point (opposite to each other).
Overlaid with a third pile-up event, with a scale hard enough to pass thresholds
of L1 triggers in an LHC experiment and realizing that each soft SD event has
a rather large cross section, such a combination of three events can
mimic our signal.

As we indicated above, due to the non-zero pile-up studied in this analysis, it
would be inefficient to require large regions of the central detector to be
empty. Instead, we require all four jets and one lepton to be well-separated
from each other and not to be accompanied by large amounts of particles, so
we introduce a cut based on the number of particles (or tracks if we work at
detector level) {from a narrow region around the primary vertex, so called
z-vertex veto, see e.g. \cite{ATLAS-zveto}}. This way the inclusive backgrounds
are believed to be suppressed even in the presence of pile-up. For this cut we
count tracks with $p_T > 0.2$~GeV and $|\eta|<2.5$ (whose efficiencies and
resolutions are properly taken into account by providing ATLAS specifications
in the Delphes card) {if they are closer than 1~mm from the primary vertex
  in the z-coordinate.}
We count the total number of such tracks per event which are at the same time
distant from the four jets and one lepton by requiring $\Delta R_{\rm trk,j} > 0.4$ and $\Delta R_{\rm trk,l} > 0.2$. We also studied two more scenarios,
  namely
\{$0.4 < \Delta R_{\rm trk,j} < 0.8$ and $0.2 < \Delta R_{\rm trk,l} < 0.8$\} and
\{$0.4 < \Delta R_{\rm trk,j} < 1.0$ and $0.2 < \Delta R_{\rm trk,l} < 1.0$\} and
found them to be less efficient than the first one. All these baseline
cuts discussed above can then be grouped as follows \\
\begin{itemize}
\item In total at least four not-overlapping jets with $E_{\rm T,jet} > 25$~GeV
  and $|\eta_{\rm jet}| < 2.5$.
\item At least one electron or muon ($\tau$ decays included)
  with $E_{\rm T,l} > 25$~GeV and $|\eta_{\rm l}| < 2.5$ isolated from all four
  jets, $\Delta R_{\rm l,j} > 0.2$.
\item At least two b-tagged jets. A jet is b-tagged if a B-hadron (generator
   level) or a b-quark (detector level) is found inside the jet.
\item FPD acceptance $0.015 < \xi_{1,2} < 0.15$.
\item Number of tracks with $p_{\rm T,trk} > 0.2$~GeV and $|\eta_{\rm trk}| < 2.5$
  in the distance {$|z_{\rm trk}-z_{\rm vtx}|<1$~mm from the primary vertex and}
  $\Delta R_{\rm trk,j} > 0.4$ from the four jets and
  $\Delta R_{\rm trk,l} > 0.2$ from one lepton must be smaller than a given value X.
\end{itemize}
and are summarized in the Table \ref{tab:cuts}.
\begin{table}[t]
\begin{tabular}{||c||} 
\hline
Cut \\ \hline\hline
$N_{\rm jet} \ge 4~(E_T>25~{\rm GeV}, |\eta|<2.5)$ \\ \hline
$N_{e/\mu} \ge 1~(E_T>25~{\rm GeV}, |\eta|<2.5)$ \\ \hline
$\Delta R (\rm{e/\mu,jet})> 0.2$ \\ \hline
$N_{\rm b-jet} \ge 2$ \\ \hline
$0.015<\xi_{1,2}<0.15 $ \\ \hline
$N_{\rm trk} (p_{\rm T} > 0.2~{\rm GeV}, |\eta| < 2.5, {|\Delta z|<1~{\rm mm}}) \leq X $\\ \hline
\end{tabular}
\caption{Cuts used in this analysis.}
\label{tab:cuts}
\end{table}


\section{Results}\label{sec:results}
\subsection{Zero pile-up scenario}
Due to much lower cross sections of the two irreducible background processes
studied in this analysis, both are considered only at zero pile-up where the
inclusive background is negligible.

The production cross section of the QED exclusive background with a $WW$ pair
in the final state at $\sqrt s = 13$~TeV is 75.6~fb, which reduces to
34~fb when taking only cases where both W-bosons decay hadronically, and
further to 4.7~fb when applying the jet $E_T$ and $\eta$ cuts. Understandably,
the most suppressing cut is the requirement of at least one lepton with
relatively high $E_T$ and which is well-isolated from all four jets. The
suppression factor of about 80 comes from the fact that leptons can only
originate from semi-leptonic decays of heavy mesons (e.g.
$D^+\to\mu^+\nu\pi^0$) and of kaons and pions inside jets and that this
occurrence naturally drops with increasing lepton $E_T$. The lepton isolation
criterion brings an additional suppression by a factor of 20 and by requiring at
least two b-tagged jets, we suppress the contribution of this background to a
negligible level (see Table~\ref{tab:cutflow_mu0}).

The photoproduction cross section of the single top quark in diffractive interactions,
$\gamma I\!\!P \rightarrow Wt$, is 12~fb as obtained from MadGraph~5, and drops by a
factor of 30 if the cuts on four jets and one lepton are applied. Another factor
of 4 comes from requiring both intact protons to be found in the FPD acceptance.
An effective cross section is of the order of 0.1~fb which is about 30 times
smaller than the effective signal cross section and hence considered to be
negligible (see Table~\ref{tab:cutflow_mu0}).

\begin{figure}[hbt!]
\includegraphics[width=0.85\textwidth, height=10cm]{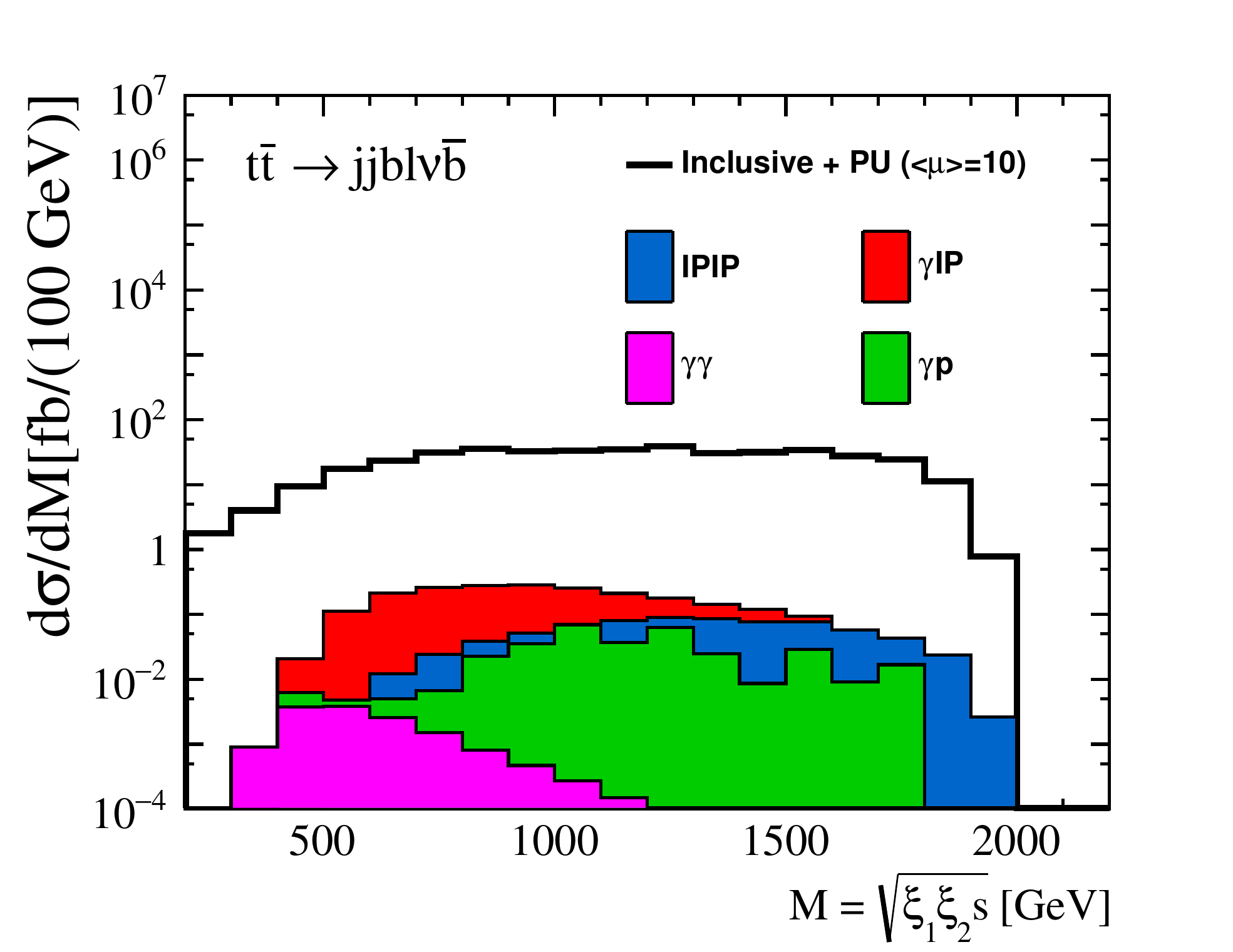}
\caption{Distribution of missing mass calculated using protons detected in FPDs
  at generator level after applying cuts in Table~\ref{tab:cuts}, except for the
  $N_{\rm trk}$ cut. Predictions for three (semi)-exclusive signal processes are
  obtained with FPMC, predictions for the $\gamma\mathbb P$ background by
   MadGraph~5, all without pile-up, while the inclusive $t\bar{t}$ background was generated with MadGraph~5+PYTHIA~8 and overlaid with pile-up with $\langle \mu \rangle = 10$ interactions per event.}
\label{fig1} 
\end{figure}

\begin{figure}[hbt!]
\includegraphics[width=0.45\textwidth]{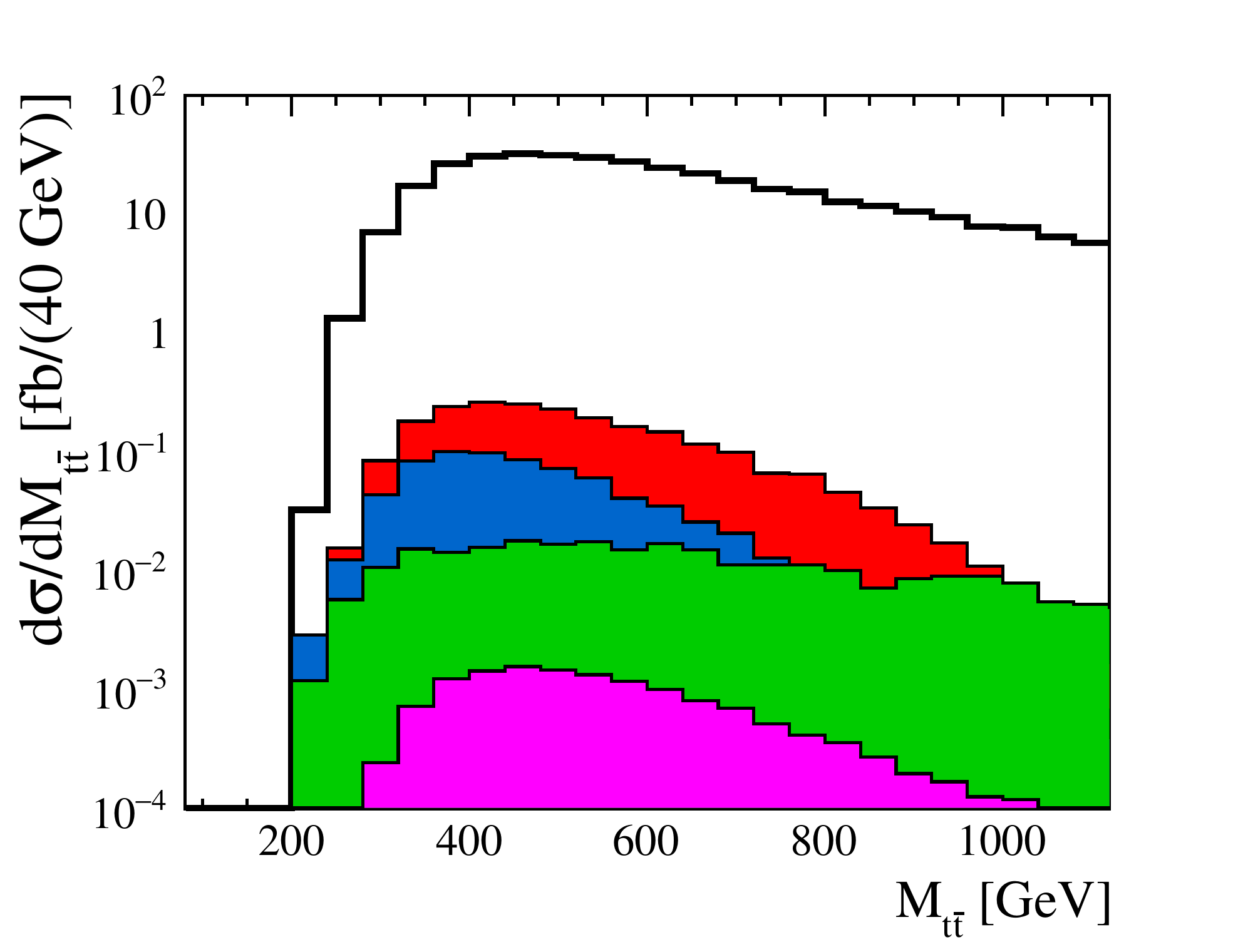}     
 \includegraphics[width=0.45\textwidth]{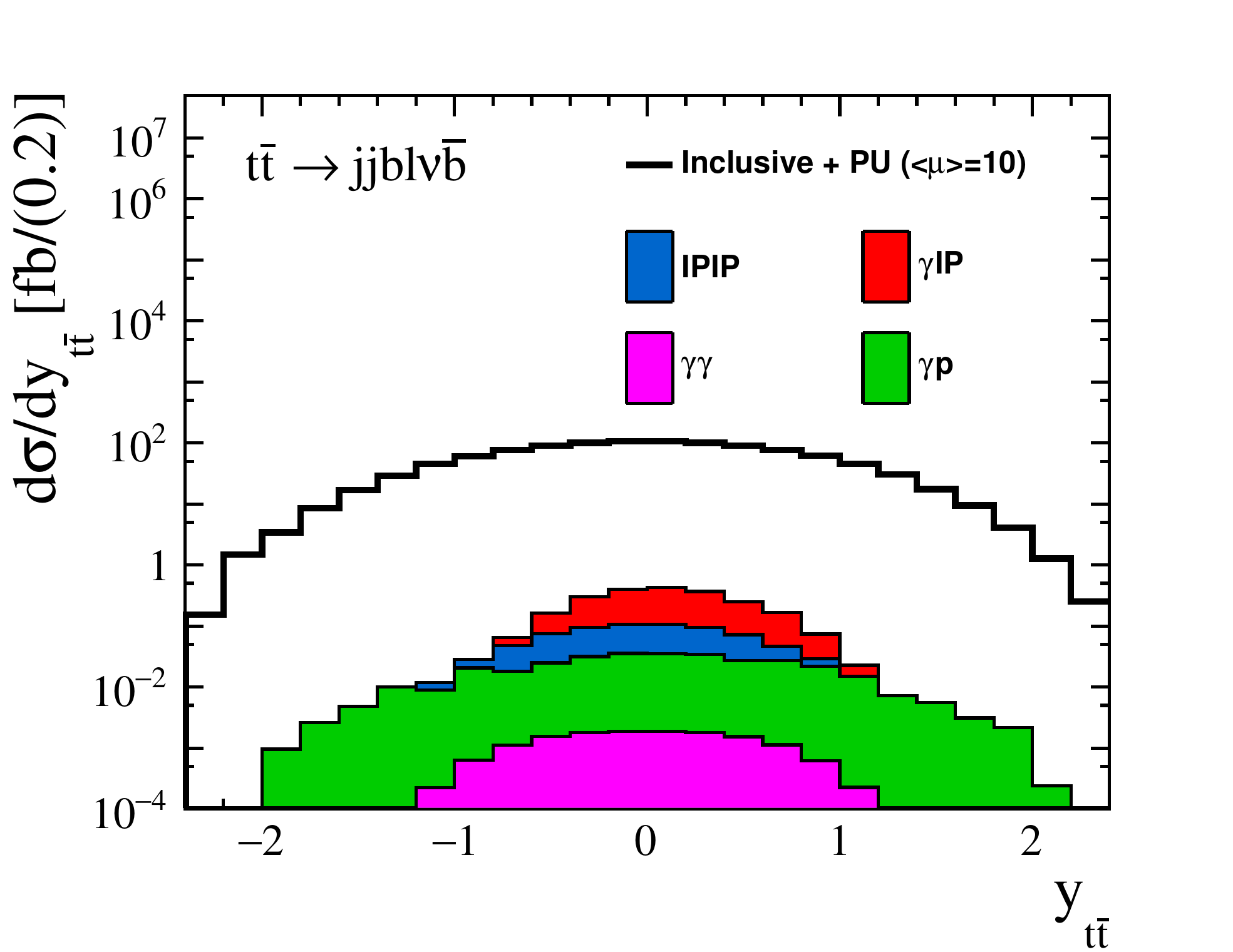}
\caption{Distribution of the mass (left) and pseudorapidity (right) of the
   $t\bar{t}$ pair at detector level and after applying cuts in
   Table~\ref{tab:cuts}, except for the $N_{\rm trk}$ cut. Predictions for
   three (semi)-exclusive signal processes are obtained with FPMC,
   predictions for the $\gamma\mathbb P$ background by
   MadGraph~5, all without pile-up, while the
   inclusive $t\bar{t}$ background was generated with MadGraph~5+PYTHIA~8 and overlaid with pile-up with $\langle \mu \rangle = 10$ interactions per event.}
\label{fig2} 
\end{figure}

A cut flow table for the zero pile-up scenario is shown in
Table~\ref{tab:cutflow_mu0}.   
\begin{table}[h]
\begin{tabular}{|c|c|c|c|c|c|c|}
\hline 
Process & \gaga & \gaP & \PP & Incl.$t\bar{t}$+PU & $\gamma\gamma\rightarrow WW$ & $\gamma\mathbb P\rightarrow Wt$ \\ \hline 
Generated cross section [fb] & 0.34& 52.0 & 28.4 & 390000 & 75.6 & 12.0\\\hline 
$N_{e/\mu} \ge 1~(E_T>25~{\rm GeV}, |\eta|<2.5)$ &0.09 & 14.1 & 7.4 & 89991 & 0.06 & 2.0 \\ \hline
$N_{\rm jet} \ge 4~(E_T>25~{\rm GeV}, |\eta|<2.5)$ &0.02  &3.9 & 2.0 & 36412 & 4.7 & 0.4 \\ \hline 
$\Delta R(\rm{e/\mu,jet})> 0.2$  & 0.02 & 3.9 & 2.0 & 36412 & 0.003 & 0.4 \\\hline
$N_{\rm b-jet} \ge 2$              & 0.02 & 3.9 & 2.0 & 36412 & $10^{-4}$ & 0.4 \\ \hline
$0.015<\xi_{1,2}<0.15 $           & 0.014 & 2.3 & 0.74 & $\sim 0$ & $\sim 0$ & 0.1 \\ \hline \hline
\end{tabular}
\caption{Cut flow for the exclusive signal processes and inclusive background
  with zero pile-up. The values marked as $\sim 0$
correspond to numbers which are sufficiently below 10$^{-4}$.} 
\label{tab:cutflow_mu0}
\end{table}
In figures \ref{fig1}, \ref{fig2} and \ref{fig3} we show control plots of some
of variables used in the event selection. All are obtained after applying cuts
in Table~\ref{tab:cuts}, except for the $N_{\rm trk}$ cut. The distributions
for the \gaga, \gaP\ and \PP\ processes are obtained from FPMC, and the
inclusive $t\bar{t}$ process from MadGraph~5+PYTHIA~8. All distributions except
for the missing mass in Fig.~\ref{fig1} are at detector level.

In Fig.~\ref{fig1} and \ref{fig2} for signal processes, no pile-up events are
added. For illustration of the combinatorial background coming from pile-up
protons, we also show the inclusive $t\bar{t}$ background with on average 10
pile-up events per interaction. The size of this combinatorial background
depends on the cross-section of the hard-scale background and the amount of
pile-up (the inclusive $t\bar{t}$ process and $\langle \mu \rangle = 10$ in this case).
In Fig.~\ref{fig1} where we plot the missing mass obtained from the $\xi$
information at generator level, we see that the most prominent background has a
shape of continuum. In Fig.~\ref{fig2} we show distributions
of the mass (left) and pseudorapidity (right) of the $t\bar{t}$ pair.
Already at this stage of analysis, we can conclude that the yield of the
exclusive \gaga\ process is negligible. The irreducible backgrounds can be
tamed to a 3\% of the signal or lower if more and better tailored cuts would be
applied. 

Running at a very low or zero amount of pile-up interactions per bunch crossing,
for example at $\mu \lesssim 1$, has a clear advantage that the combinatorial
background overlaid with a hard-scale inclusive background processes with large
cross sections, would become negligible. However, conceivable values of
integrated luminosity (they would be smaller than 1~fb$^{-1}$) do not allow one
to imagine measuring the top quark mass reliably since it would be based on
fewer than 10~events in the whole reachable range between $2\cdot m_t$ and
roughly 2.5~TeV. Relaxing the $p_T$ cuts for leptons (down to 5~GeV) and jets
(down to 20 GeV) does not help since it would increase the signal statistics by
only about 25\%.

\subsection{Non-zero pile-up scenario}
From figures Fig.~\ref{fig1} and \ref{fig2} we observe that the contamination by
the mix of inclusive and combinatorial backgrounds is enormous and needs to be
suppressed by special means. As discussed above, these are referred to as the
exclusivity requirement and the suppression from using the ToF detector,
and are discussed in this section.

The hard-scale process for the inclusive $t\bar{t}$ production is generated
using MadGraph5~\cite{madgraph}, while showering and hadronization is done by
PYTHIA~8 \cite{pythia}. The Delphes package is then used for two purposes:
i) pile-up mixing and ii) inclusion of detector effects. First, it overlays the
inclusive (hard-scale) $t\bar{t}$ events with pile-up
(usually of soft nature), in other words it mixes one hard-scale event with a
given number of pile-up events such that the resulting distribution of number
of pile-up events when integrated over all hard-scale events, follows a
Poissonian distributions with a mean equal to $\langle \mu \rangle$. We study
in detail three working points, namely $\langle \mu \rangle$ of 5, 10 and 50.
The pile-up events are generated by PYTHIA~8 as minimum-bias events at 13~TeV
with default settings, i.e. multi-parton interactions (MPI), initial as well as
final state radiations (ISR and FSR) are all switched on. Based on the knowledge
of a probability to see an intact proton from one minimum bias event in the FPD
$\xi$-acceptance on one side, $P_{\rm ST}$, we are able to estimate the rate of
fake double-tagged events and hence the combinatorial background from pile-up.
For the minimum bias events generated as specified above, we get $P_{\rm ST} = 1.4$\%. The combinatorial factors representing the rates of fake double-tagged
events for the studied values of $\langle \mu \rangle$ of 5, 10 and 50 are
{0.0031}, 0.014 and 0.246, respectively (see \cite{Tasevsky:2014cpa,Harland-Lang:2018hmi,ToFperformance} for more details and the $\langle \mu \rangle$-dependence of this
background). 

Second, Delphes provides a fast simulation of all relevant detector features
for which parameters are put in the input card. We used those specific for the
ATLAS detector.

\begin{figure}[htbp!]
\includegraphics[width=0.325\textwidth,height=4.7cm]{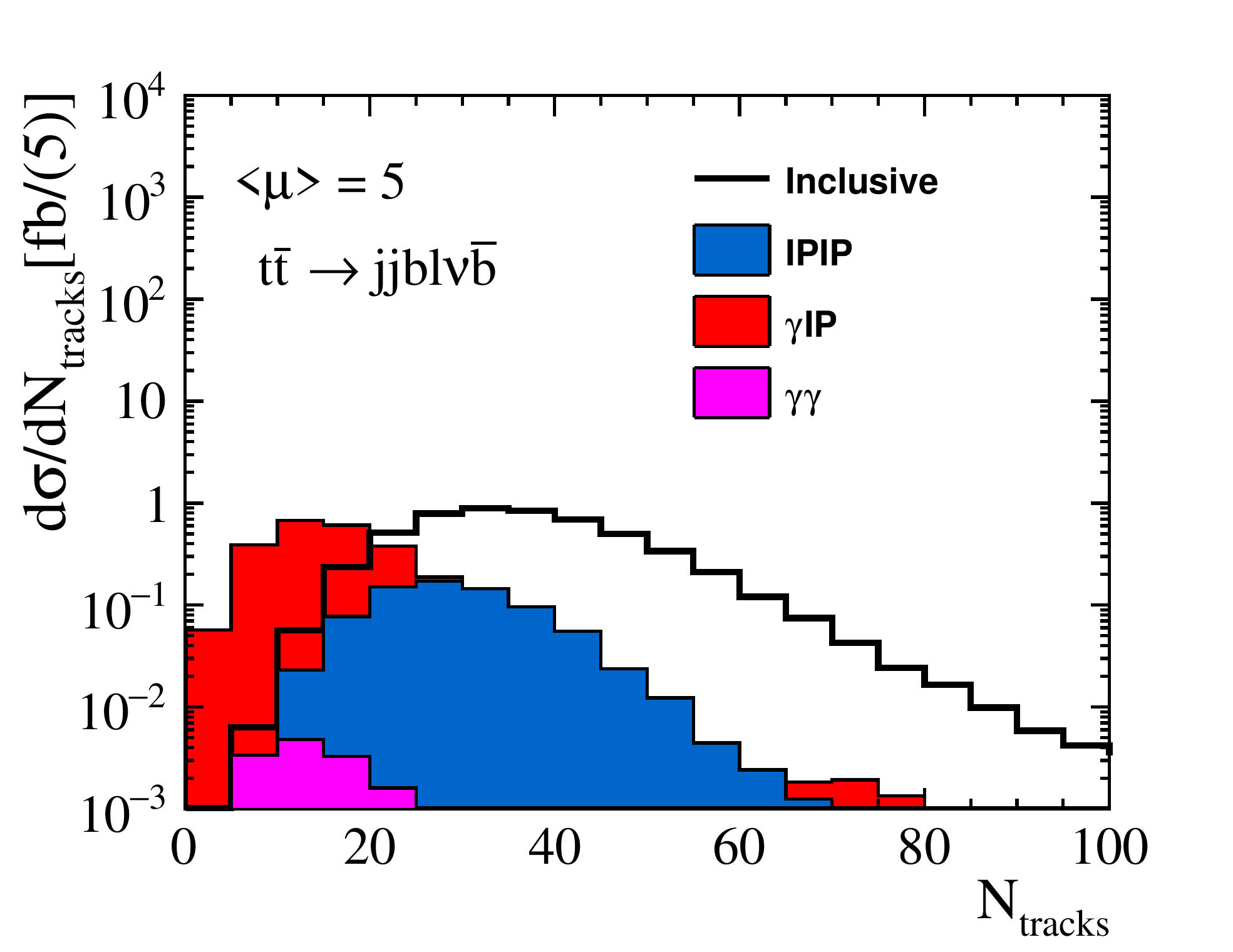}
\includegraphics[width=0.325\textwidth,height=4.7cm]{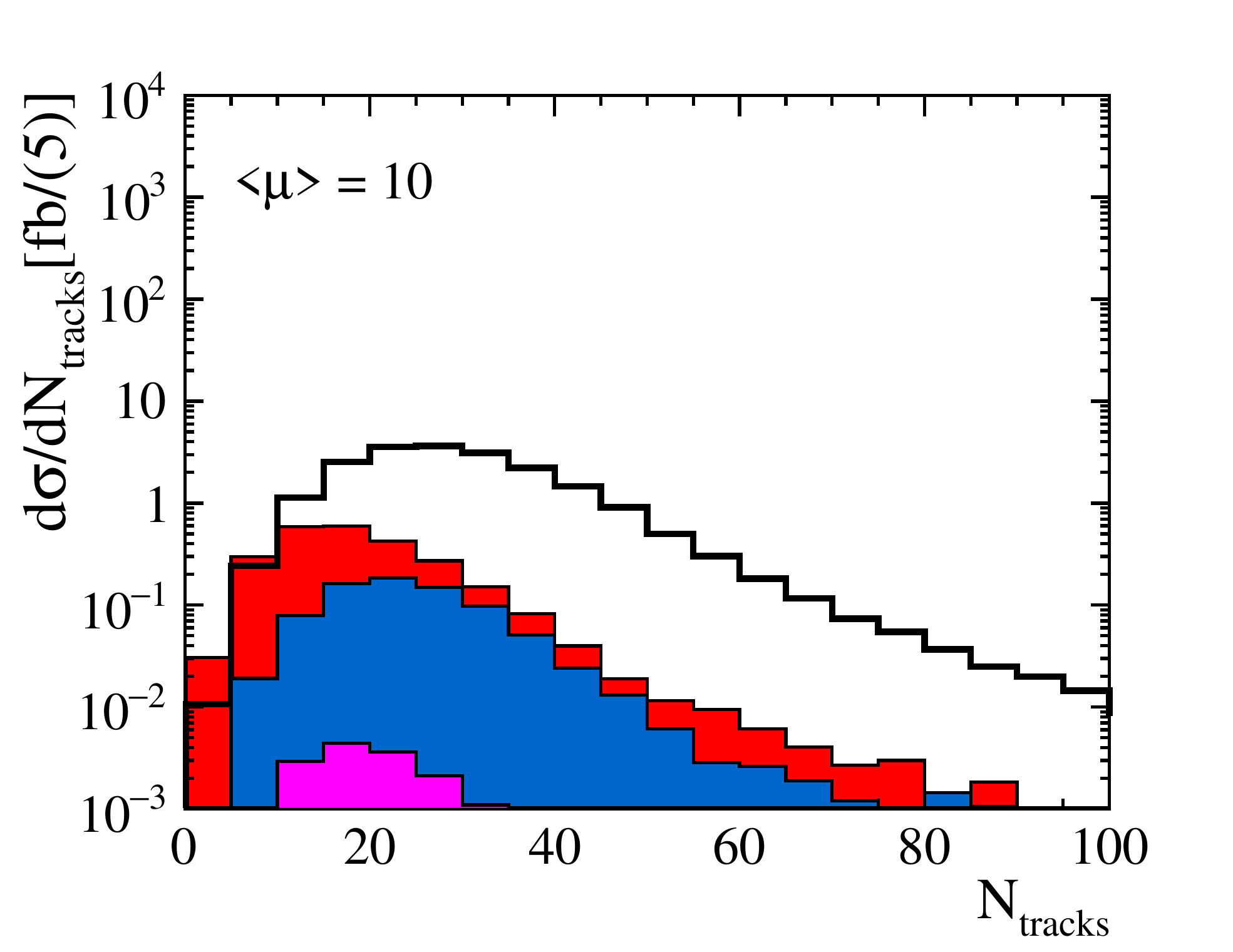}
\includegraphics[width=0.325\textwidth,height=4.7cm]{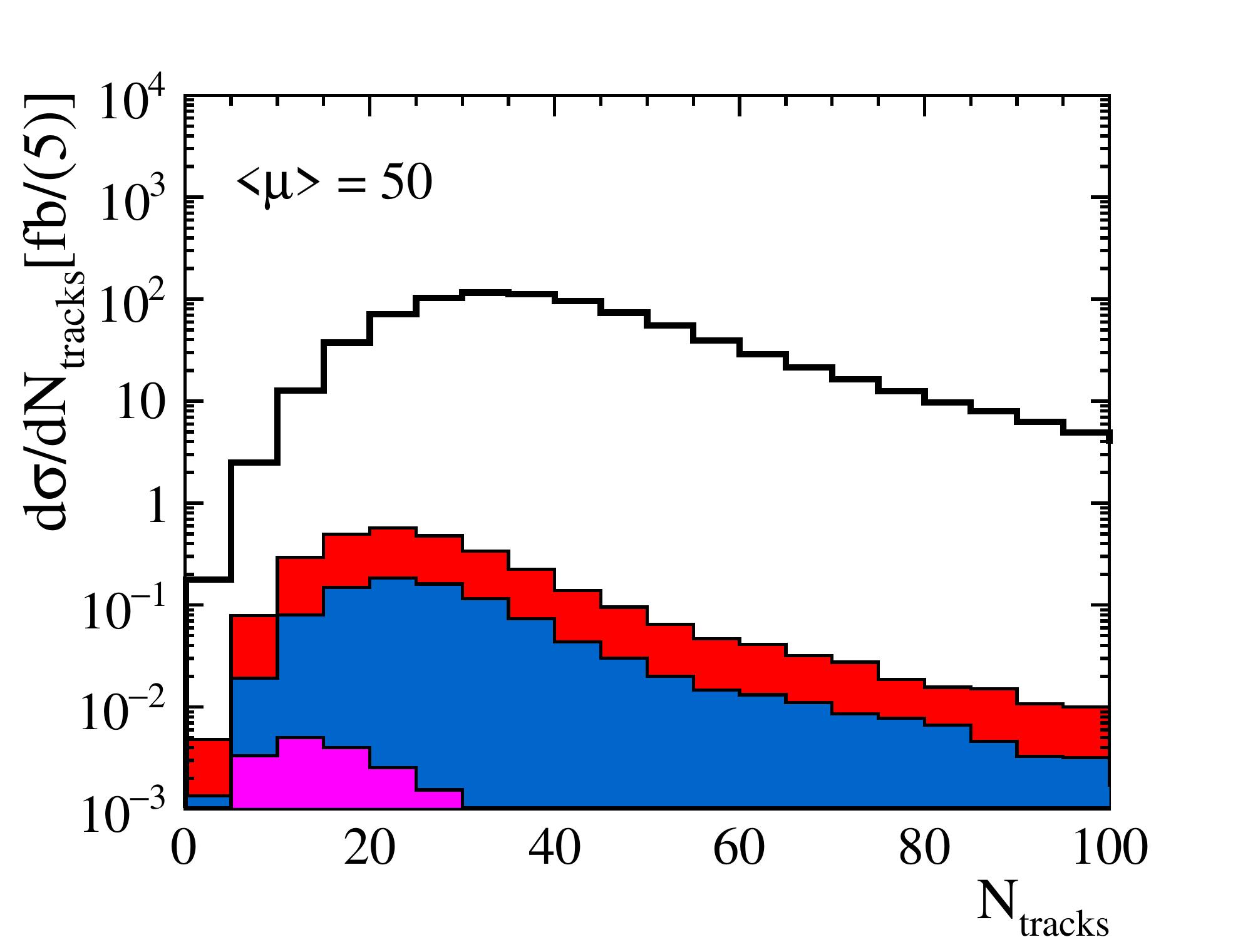}
\caption{Distribution of the number of tracks with $p_T > 0.2$~GeV and
  $|\eta| < 2.5$ outside all four jets and one lepton for three amounts of
  pile-up events per interaction, $\langle \mu \rangle$, of 5, 10 and 50, all
  at detector level and after applying cuts in Table~\ref{tab:cuts}, except for
  the $N_{\rm trk}$ cut. Predictions for three (semi)-exclusive signal processes
  are obtained with FPMC, while the inclusive $t\bar{t}$ background was
  generated with MadGraph~5+PYTHIA~8.}
\label{fig3}
\end{figure}

The distribution of number of tracks outside the four jets and one lepton for
three amounts of pile-up, namely for $\langle \mu \rangle =$~5, 10 and 50, is
shown in Fig.~\ref{fig3}. Tracks are required to have $p_T > 0.2$~GeV and
be in the central tracker acceptance, $|\eta| <$~2.5. When plotting these
distributions, the effective cross sections of the mix of inclusive and pile-up
events are already scaled by the corresponding rates of fake double tagged
events (specified above) and by ToF suppression factors. The ToF suppression
factors are 18.3, 17.3 and 10.8 for $\langle \mu \rangle =$~5, 10 and 50,
respectively, under the assumption that the time resolution of ToF is
$\sigma_t =$~10~ps and the signal is collected in a 2-$\sigma_t$ window. The
effective cross-sections after applying individual cuts from
Table~\ref{tab:cuts} and scaling by the rates of fake double tagged events and
by the ToF suppression, are summarized in Table~\ref{tab:cutflow_muall}. To
suppress the dominant background further, we also added a cut on the missing
mass evaluated by FPD and the mass of the top pair measured in the central
detector. 

\begin{table}[h]
\begin{tabular}{|c|c|c|c|}
\hline 
Process & \gaP($\langle \mu \rangle$=5/10/50) & \PP($\langle \mu \rangle$=5/10/50) & Incl.$t\bar{t}$+PU($\langle \mu \rangle$=5/10/50) \\ \hline 
Generated cross section [fb] & 52.0 & 28.4 & 390000 \\\hline 
$N_{e/\mu} \ge 1~(E_T>25~{\rm GeV}, |\eta|<2.5)$ & 14.1/14.2/13.4 & 7.4/7.3/6.7 & 90057/90042/82994 \\ \hline
$N_{\rm jet} \ge 4~(E_T>25~{\rm GeV}, |\eta|<2.5)$ &4.2/4.4/5.4 & 2.1/2.2/2.6 & 38157/38928/42821 \\\hline 
$\Delta R(\rm{e/\mu,jet})> 0.2$  & 4.2/4.4/5.4 & 2.1/2.2/2.6 & 38157/38928/42821 \\\hline
$N_{\rm b-jet} \ge 2$              & 4.2/4.4/5.4 & 2.1/2.2/2.6 & 38157/38928/42821 \\ \hline
$0.015<\xi_{1,2}<0.15 $           & 2.4/2.6/3.2 & 0.8/0.8/1.0 & 118.2/423.3/10534 \\ \hline
$m_{t\bar{t}}<1000$~GeV, $m_X>400$~GeV & 2.4/2.6/3.1 & 0.8/0.8/1.0 & 97.6/349.6/9107 \\ \hline
ToF suppression & 2.4/2.6/2.4 & 0.8/0.8/0.8 & 5.3/20.2/843.2 \\ \hline
$N_{\rm trk} \leq 10 $ & 0.45/0.44/0.14 & 0.002/0.02/0.02 & 0.006/0.35/2.7\\ \hline
$N_{\rm trk} \leq 15 $ & 1.12/1.12/0.60 & 0.10/0.10/0.10 & 0.12/1.39/15.4\\ \hline
$N_{\rm trk} \leq 20 $ & 1.73/1.76/1.20 & 0.11/0.26/0.25 & 0.29/3.94/52.8\\ \hline
$N_{\rm trk} \leq 25 $ & 2.11/2.16/1.80 & 0.30/0.45/0.44 & 0.81/7.49/123.9\\ \hline
\end{tabular}
\caption{Cut flow for the effective cross sections in femtobarns for the
  exclusive signal processes and inclusive background with pile-up overlaid
  with $\langle \mu \rangle$ = 5, 10 and 50. The effect of the $\xi$ cut for
  the inclusive background with pile-up is evaluated as a combinatorial
  background coming from the rate of fake double-tagged events. Suppression of
  pile-up effects from using ToF information is based on \cite{Tasevsky:2014cpa,Harland-Lang:2018hmi}.}
\label{tab:cutflow_muall}
\end{table}

To estimate the statistical significance, $\sigma$, and signal to background
ratio, S/B, we consider three luminosity scenarios in terms of
$\langle \mu \rangle$ and $\cal L$ where $\langle \mu \rangle$ represents the
average number of pile-up interactions per event (or the instantaneous
luminosity) and $\cal L$ is the integrated luminosity. We assume
$\cal L$ to be 10, 30 and 300~fb$^{-1}$ for $\langle \mu \rangle =$~5, 10 and
50, respectively, and provide the information on above for four $N_{\rm trk}$
values in Table~\ref{tab:sig}. The best combination of $\sigma$ and S/B parameters give track cuts
$N_{\rm trk} \leq 15$ or $N_{\rm trk} \leq 20$ for all luminosity scenarios with
significances about 11, 6 and 3 going from the lowest to the largest luminosity
scenario. {{These significances are rather insensitive to non-negligible
    uncertainties connected with the chosen value of the $S^2$ factor.
    In most cases, differences are below 5\% and safely below 10\% in the rest
    when changing the $S^2$ value from 3\% to 2\% or 4\%.}

\begin{table}[h]
\begin{tabular}{|c|c|c|c|}
\hline 
($\langle \mu \rangle, \cal L$[fb$^{-1}$]) & (5, 10) & (10, 30) & (50, 300) \\ \hline 

$N_{\rm trk} \leq 10$ & 4.52/0.06, 18.5 & 13.8/10.5, 4.3 & 48.3/810.0, 1.7\\\hline
$N_{\rm trk} \leq 15$ & 12.2/1.2,  11.1 & 36.6/41.7, 5.7 & 195/4616, 2.9 \\\hline
$N_{\rm trk} \leq 20$ & 18.3/2.9,  10.7 & 60.6/118.2, 5.6& 429/15827, 3.4\\\hline
$N_{\rm trk} \leq 25$ & 23.6/8.1,  8.3  & 78.3/224.7, 5.2& 672/37195, 3.5\\\hline
\end{tabular}
\caption{Summary of event yields for four values of $N_{\rm trk}$ cut and for
  three luminosity scenarios ($\langle \mu \rangle$, $\cal L$) where $\cal L$
  stands for integrated luminosity in fb$^{-1}$.
  For each scenario, the ratio of signal to background events,
  $N_{\rm S}/N_{\rm B}$, and a statistical significance is given.}
\label{tab:sig}
\end{table}

For a sensible measurement of the top quark mass, one would need a sufficient
amount of signal events and a very low level of background contamination.
As an example, we took the missing mass spectrum generated with top quark mass
of 171.7~GeV (about $4\,\sigma$ from the current top mass world average
172.8~GeV) and calculated the value of $\chi^2$, assuming the template is
modeled by a top quark mass of 172.8~GeV. For one of the best configurations
of those studied in this analysis, namely for ($\langle \mu \rangle, \cal L$) =
(5, 10), the differences are much below $1\,\sigma$ (see the corresponding
missing mass spectrum in Fig.~\ref{missmass_5-10} with signal and inclusive
background after applying the $N_{\rm trk} \leq 20$ cut).

\begin{figure}[hbt!]
\includegraphics[width=0.85\textwidth, height=10cm]{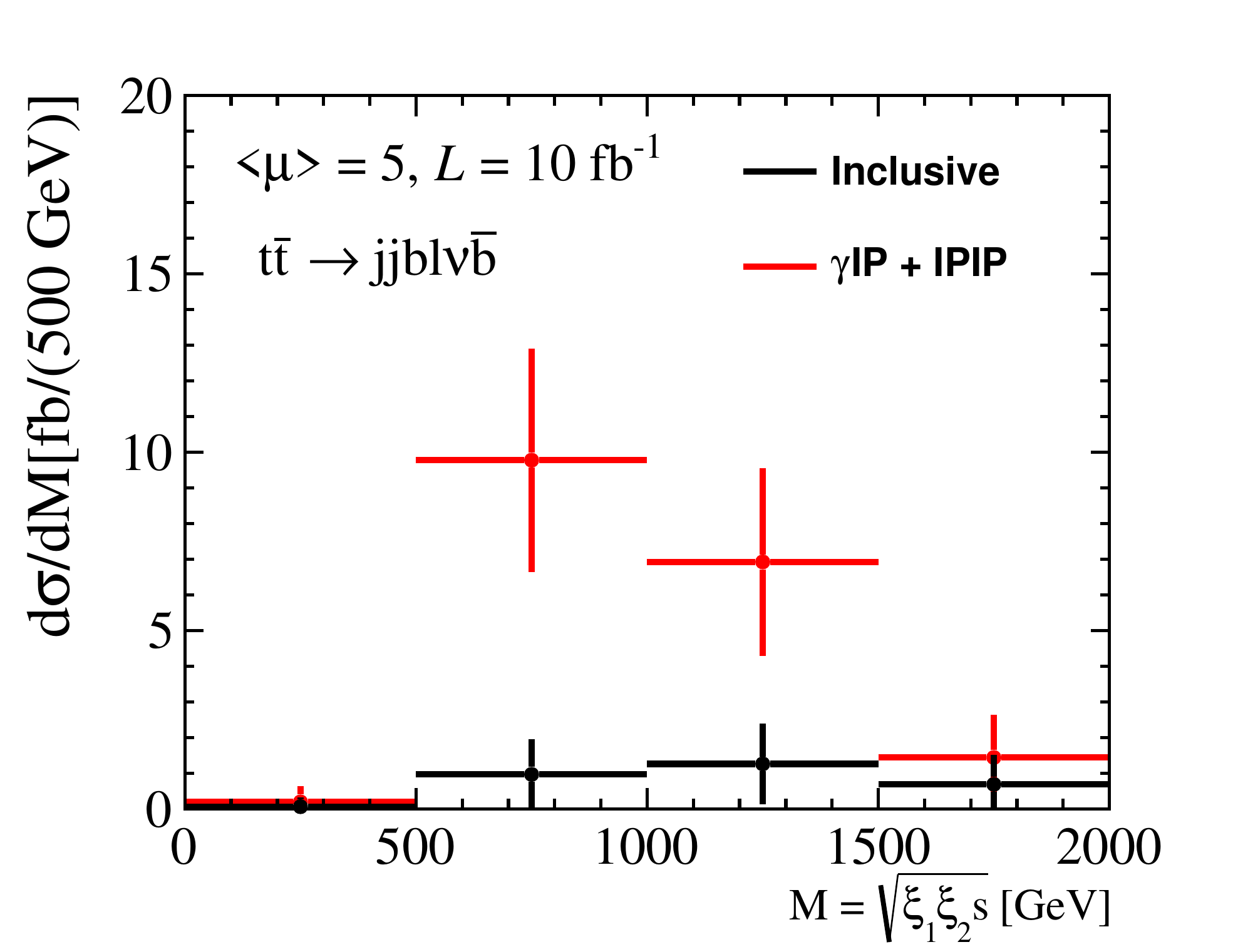}
\caption{Distribution of missing mass calculated using protons detected
  in FPDs at generator level after applying cuts in Table~\ref{tab:cuts}
  {and corresponding ToF suppression} and $N_{\rm trk} \leq 20$ cut. Predictions for the two semi-exclusive signal
   processes are obtained with FPMC, while the inclusive $t\bar{t}$ background
   was generated with MadGraph~5+PYTHIA~8. All are overlaid with pile-up with
   $\langle \mu \rangle = 5$ interactions per event and numbers of events
   correspond to the integrated luminosity of 10~fb$^{-1}$.}
\label{missmass_5-10} 
\end{figure}
When we enlarge statistics
150 times, the p-values are 0.78, 0.24 and 0.42 for $N_{\rm trk} \leq 15, 20$ and
25, respectively. So only for the $N_{\rm trk} \leq 20$ case we start to see
a $1\,\sigma$ effect. In other words in order to distinguish a 1.2~GeV
difference in measured top quark mass at a $1\,\sigma$ significance (considering
statistical uncertainties only), a sample of 1500~fb$^{-1}$ collected at
$\langle \mu \rangle = 5$ would be needed. In other studied luminosity and
$N_{\rm trk}$ points, the situation is even worse. Therefore we conclude that  
overall the situation with S/B values does not give favourable prospects for
measuring precisely the mass of the top quark. Let us note in this context that
even lowering the ToF resolution to 5~ps
would only lead to halving the combinatorial background which is
clearly not sufficient from the statistical point of view. 

\section{Summary}
\label{sec:sum}

We studied in detail prospects for measuring the $t\bar{t}$ pair produced in the
exclusive (\gaga) and semi-exclusive (\gaP\ and \PP) processes. We analyzed four
luminosity scenarios, going from zero pile-up up to $\langle \mu \rangle$ of 50
with corresponding assumed integral luminosities of up to 300~fb$^{-1}$. With
the help of Delphes, the main effects of detector acceptance and resolutions
as well as the effect of pile-up background were included in the analysis
procedure. Good prospects for observing the exclusive signal over a mixture of
inclusive and combinatorial background are achieved for all luminosity
scenarios, although a good separation between the two are observed for rather
low amounts of pile-up, typically lower than $\langle \mu \rangle$ of 50.
Statistical significances evaluated from estimated numbers of signal and
background events are around 3 for the highest luminosity scenario
($\langle \mu \rangle, \cal L$[fb$^{-1}$]) = (50, 300), about 6 for the (10,30) and 11
for the (5,10) scenarios. From a simple statistical analysis, we find that
these significances are still not sufficient for a determination
of the top quark mass that would be competitive with inclusive methods.
Much higher statistics would be needed with the current experimental procedure
or more sophisticated procedures to suppress the dominant background have to
be developed. 


\begin{acknowledgments}
The authors thank James Howarth for the initial inspiration of this study. VPG thank the members of the Institute of Physics of the Czech Academy of Science by the warm hospitality during the beginning  of this project.  
This work was partially financed by the Brazilian funding agencies CNPq, CAPES,
FAPERGS, FAPERJ and INCT-FNA (processes number 464898/2014-5 and
88887.461636/2019-00). MT is supported by MEYS of the Czech Republic within
project LTT17018.
\end{acknowledgments}

\end{document}